\documentclass[aps,twocolumn,showpacs,superscriptaddress,amssymb]{revtex4}

\usepackage{graphicx,amssymb}

\begin{document}

\title{Understanding the different rotational behaviors of $^{252}$No and $^{254}$No}

\author{H. L. Liu}
\email{hlliu@mail.xjtu.edu.cn} \affiliation{Department of Applied
Physics, Xi'an Jiaotong University, Xi'an 710049, China}
\author{F. R. Xu}
\affiliation{School of Physics, Peking University, Beijing 100871,
China}
\author{P. M. Walker}
\affiliation{Department of Physics, University of Surrey,
Guildford, Surrey GU2 7XH, UK} \affiliation{CERN, CH-1211 Geneva
23, Switzerland}

\begin{abstract}
Total Routhian surface calculations have been performed to
investigate rapidly rotating transfermium nuclei, the heaviest
nuclei accessible by detailed spectroscopy experiments. The
observed fast alignment in $^{252}$No and slow alignment in
$^{254}$No are well reproduced by the calculations incorporating
high-order deformations. The different rotational behaviors of
$^{252}$No and $^{254}$No can be understood for the first time in
terms of $\beta_6$ deformation that decreases the energies of the
$\nu j_{15/2}$ intruder orbitals below the $N=152$ gap. Our
investigations reveal the importance of high-order deformation in
describing not only the multi-quasiparticle states but also the
rotational spectra, both providing probes of the single-particle
structure concerning the expected doubly-magic superheavy nuclei.
\end{abstract}

\pacs{21.10.Re, 21.60.Cs, 23.20.Lv, 27.90.+b}

\maketitle

Together with the exploration of nuclei far away from the
stability line using radioactive nuclear beams, the synthesis of
superheavy nuclei towards the predicted ``island of stability'' by
fusion reactions is the focus of current research on atomic
nuclei~\cite{HofRMP00,OgaJPG07}. The occurrence of superheavy nuclei
owes to the quantum shell effect (see, e.g., Ref.~\cite{BendPLB01}) that overcomes the
strong Coulomb repulsion between the large number of protons.
The shell effect peaks at the expected doubly-magic nucleus
next after $^{208}$Pb, the center of the stability island.
Unfortunately, various theories give rise to different magic
numbers and available experiments have not been able to confirm or
exclude any of them. Nevertheless, one can obtain single-particle
information that is intimately related to the shell structure of
superheavy nuclei from transfermium nuclei where $\gamma$-ray
spectroscopy has been accessible for
experiments~\cite{HerzPPNP08,HerzActa11}.

Transfermium nuclei have been found to be deformed. For example,
$\beta_2\approx 0.27$ has been derived for $^{254}$No from the
measured ground-state band~\cite{ReitPRL99}. The observation of
$K$ isomers with highly-hindered decays in
$^{254}$No~\cite{HerzNature06,TandPRL06,ClarkPLB10} points to an
axially-symmetric shape for the nucleus. The deformation can bring
the single-particle levels from the next shell across the
predicted closure down to the Fermi surface. They play an active
role in both nuclear non-collective and collective motions that in
turn can serve as probes of the single-particle structure. For
example, the observed $K^\pi=3^+$ state formed by broken-pair
excitation in $^{254}$No is of special
interest~\cite{HerzNature06}. This is because the $\pi1/2^-[521]$
orbital occupied by one unpaired nucleon stems from the spherical
orbital $2f_{5/2}$ whose position relative to the spin-orbit
partner $2f_{7/2}$ determines whether $Z=114$ is a magic number
for the ``island of stability''. On the other hand, high-$j$
intruder orbitals sensitively respond to the Coriolis force during
collective rotation. The observation of upbending or backbending
phenomena is usually associated with the alignment of high-$j$
intruder orbitals. The spectroscopy experiments on transfermium
nuclei provide a testing ground for the theoretical models that
are used to predict the properties of superheavy nuclei.

\begin{figure}[b]
\includegraphics[scale=0.45]{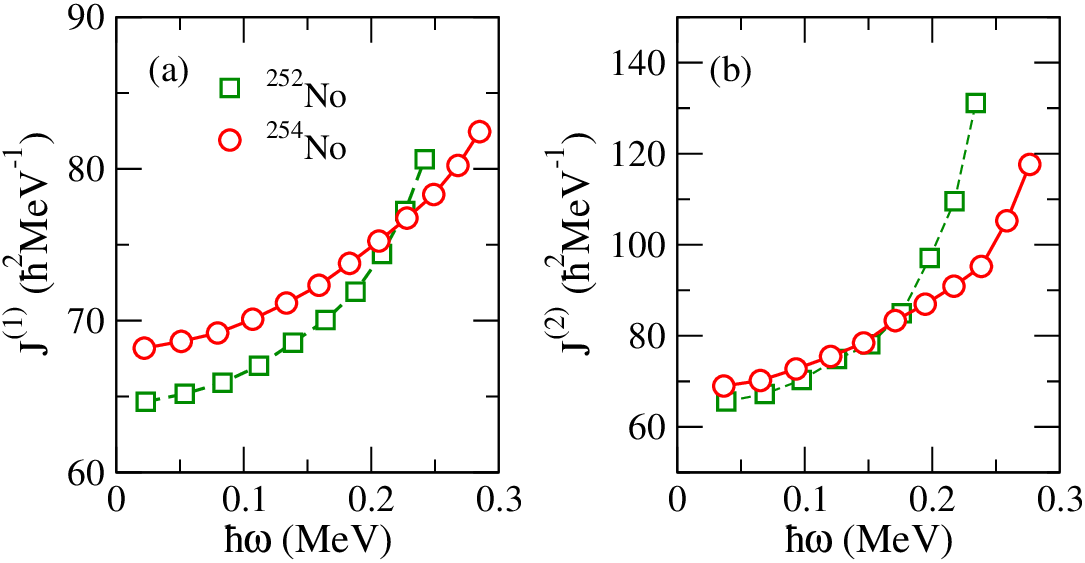}
\caption{\label{fig1}(Color online) Experimental kinematic (a) and
dynamic (b) moments of inertia for $^{252,254}$No. Data are taken
from Refs.~\cite{HerzPRC01} and \cite{EktEPJA05} for $^{252}$No
and $^{254}$No, respectively.}
\end{figure}

\begin{figure*}
\includegraphics[scale=0.6]{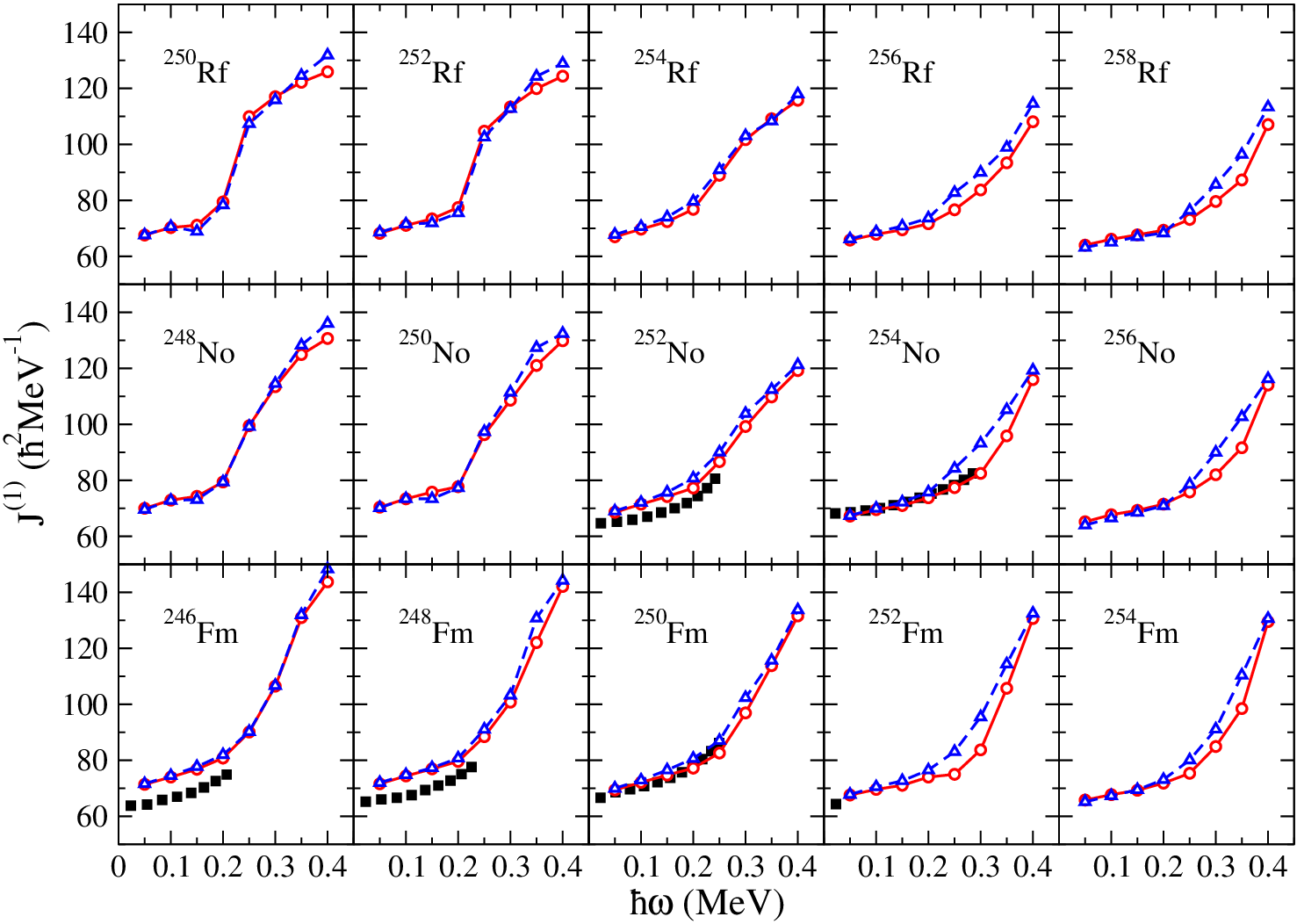}
\caption{\label{fig2}(Color online) Calculated kinematic moments
of inertia for Fm, No, and Rf isotopes, compared with available
experimental
data~\cite{HerzPPNP08,HerzActa11,ReitPRL99,HerzPRC01,EktEPJA05,PioPRC12,BasPRC06,GrePRC08}.
Calculations with and without high-order deformations
$\beta_{6,8}$ are represented by open circles (red) and open
triangles (blue), respectively. Filled squares (black) indicate
experimental data.}
\end{figure*}

Modern in-beam spectroscopy experiments have observed rotational
bands up to high spins, such as in the even-even nuclei
$^{246}$Fm~\cite{PioPRC12}, $^{248}$Fm~\cite{HerzPPNP08},
$^{250}$Fm~\cite{BasPRC06,GrePRC08}, $^{252}$No~\cite{HerzPRC01},
and $^{254}$No~\cite{ReitPRL99,EktEPJA05}. Especially, the
yrast spectrum of $^{254}$No has been extended to spins of more
than 20$\hbar$ because of the relatively high production rate.
Theoretically, various models have been applied to study the
rotational properties of transfermium nuclei. The calculations
include: (i) cranking approximations of mean-field models such as
the macroscopic-microscopic
approach~\cite{MuntPRC99,MuntPLB01,SobPRC01}, the Nilsson
potential with the particle-number-conserving
method~\cite{HeNPA09,ZhPRC11,ZhPRC12}, the Hartree-Fork-Bogoliubov
(HFB) approach with the Skyrme force~\cite{DuguNPA01,BendNPA03},
the HFB approach with the Gogny force~\cite{EgidPRL00,DelaNPA06},
and the relativistic Hartree-Bogoliubov approach~\cite{AfanPRC03};
(ii) the projected shell model~\cite{SunPRC08,ChenPRC08,AlkPRC09}
that incorporates beyond-mean-field effects, restored symmetry and
configuration mixing. In general, the theories can reproduce the
observations.

However, it is still an open question why $^{252}$No and
$^{254}$No exhibit significantly different rotational behavior at
high spins. The difference has been noticed since 2001 when the
rotational band up to $I=20$ in $^{252}$No was first
observed~\cite{HerzPRC01}. In Fig.~\ref{fig1}, the experimental
moments of inertia (MOIs) are displayed for $^{252,254}$No. It is
seen that the MOIs of both nuclei increase gradually with
rotational frequency at low spins. When reaching
$\hbar\omega\approx 0.2$ MeV, the $^{252}$No MOI grows sharply,
while the $^{254}$No MOI remains steady until a rotational
frequency approaching 0.3 MeV. Although the previous various
calculations can reproduce the MOIs of $^{252}$No and $^{254}$No, no one
has explained in detail the mechanism responsible for the significant
MOI difference of the two nuclei. Using total Routhian surface (TRS)
calculations, now extended to include high-order deformations, we
show that the difference can be understood in terms of
$\beta_6$ that decreases the energies of the
$\nu j_{15/2}$ intruder orbitals below the $N=152$ deformed shell gap.
We note that Schunck {\it et al.}~\cite{DudkPRC07}
performed TRS calculations with high-order deformations. However,
they did not include pairing correlations, which are necessary for
the description of upbending and backbending phenomena.

The TRS approach~\cite{SatuNPA94,SatuPST95,XuNPA00} adopted here
is a pairing-deformation-frequency self-consistent calculation
based on the cranked shell model. The single-particle states are
obtained from the axially deformed Woods-Saxon
potential~\cite{CwCPC87} with the parameter set widely used for
cranking calculations. Both monopole and doubly-stretched
quadrupole pairings are included. The monopole pairing strength
($G$) is determined by the average-gap method~\cite{MolNPA92}, and
the quadrupole pairing strengths are obtained by restoring the
Galilean invariance broken by the seniority pairing
force~\cite{SakPLB90}. The quadrupole pairing has negligible
effect on energies, but it is important for the proper description
of MOIs~\cite{SatuPST95}. An approximate particle-number
projection is carried out by means of the Lipkin-Nogami
method~\cite{LipNPA73}, thus avoiding the spurious collapse of
pairing correlations at high angular momentum. For any given
deformation and rotational frequency, the pairings are
self-consistently calculated by the HFB-like
method~\cite{SatuNPA94}, so the dependence of pairing correlations
on deformation and frequency is properly treated. The total energy
of a state consists of a macroscopic part that is obtained with
the standard liquid-drop model~\cite{MyeNP66}, a microscopic part
that is calculated by the Strutinsky shell-correction
approach~\cite{StruNPA67}, and the contribution due to rotation.
At each frequency, the deformation of a state is determined by
minimizing the TRS calculated in a multi-dimensional deformation
space.

In the present work, the deformation space includes $\beta_2$,
$\beta_4$, $\beta_6$, and $\beta_8$ degrees of freedom.
Transfermium nuclei around $^{254}$No have been predicted to have
$\beta_6$ deformations~\cite{MuntPLB01,SobPRC01}. It has been
demonstrated that the $\beta_6$ deformation leads to enhanced
deformed shell gaps at $Z=100$ and $N=152$~\cite{PatPLB91}, and
has remarkable influence on the binding energy~\cite{PatNPA91},
ground-state MOI~\cite{MuntPLB01}, and $K$-isomer excitation
energy~\cite{LiuPRC11}. With the inclusion of $\beta_{6,8}$
deformations in the TRS calculation, we investigate their effect
on the collective rotation at high spins.

Figure~\ref{fig2} displays the kinematic MOIs calculated with and
without high-order deformations for Fm, No, and Rf isotopes. All
the calculations are performed without any adjustment of the
parameters. When compared with experiments, the calculations
including $\beta_{6,8}$ deformations are in good agreement with
the data for $^{250}$Fm and $^{254}$No. In particular, we
reproduce well the slow alignment observed in $^{254}$No. Here an
upbending is predicted beyond $\hbar\omega\approx 0.3$ MeV. The
upbending at $\hbar\omega\approx 0.2$ MeV observed in $^{250}$Fm
is also reproduced well. Our calculations, however, overestimate
the measured MOIs in $^{246,248}$Fm and $^{252}$No. Nevertheless,
the theoretical variation trends are consistent with observations.
It is worth noting that the observed quick growth of MOI at
$\hbar\omega\approx 0.2$ MeV in $^{252}$No is also shown in our
calculation. For $^{246,248}$Fm, we predict the occurrence of
upbending at $\hbar\omega\approx 0.2$ MeV that is in the proximity
of the available experimental values. A few more data points will
test the predictions.

Similar overestimation of the MOI in TRS calculations has also
been seen in rare-earth nuclei, which is ascribed to the
deficiency of the monopole pairing strength given by the average
gap method~\cite{XuPRC99}. It was found that an improved pairing
strength can be obtained by matching the experimental and
theoretical odd-even mass differences. The new pairing strength,
including mean-field and blocking effects, results in a better
description of the MOI~\cite{XuPRC99} and the multi-quasiparticle
state excitation energy~\cite{XuPLB98}. Nevertheless, the
adjustment of $G$ seems to barely affect the backbending/upbending
frequency. Since the variation trends of the observed MOIs in
transfermium nuclei can be reproduced and the present focus is on
the high-spin property difference between $^{252}$No and
$^{254}$No, we do not fine tune the pairing strength in this work.

\begin{figure}
\includegraphics[scale=0.45]{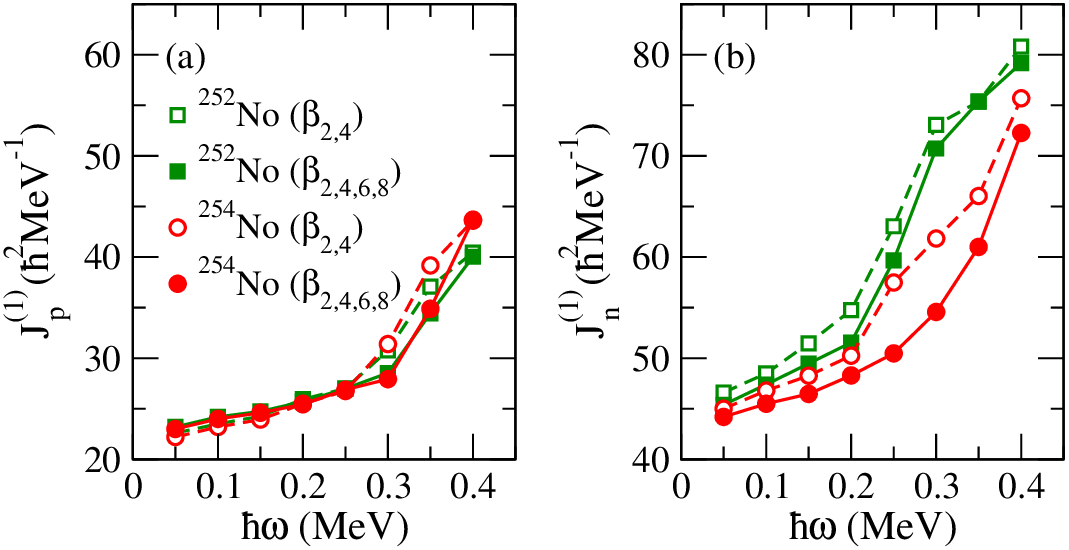}
\caption{\label{fig3}(Color online) Proton (a) and neutron (b)
components of the calculated kinematic moments of inertia for
$^{252,254}$No.}
\end{figure}

The backbending/upbending phenomenon in the MOI is usually
associated with the rotation alignment of high-$j$ intruder
orbitals near the Fermi surface. In the transfermium region, the
intruder orbitals are $\pi i_{13/2}$ for protons and $\nu
j_{15/2}$ for neutrons. The relativistic Hartree Bogliubov
model~\cite{AfanPRC03}, the cranked shell model with particle
number conservation~\cite{ZhPRC12}, and the projected shell
model~\cite{AlkPRC09} all indicated the competitive alignments of
$\pi i_{13/2}$ and $\nu j_{15/2}$ orbitals that take place
simultaneously. Our calculations show, however, that the upbending
is mostly ascribed to the alignment of the $\nu j_{15/2}$ orbital,
with some contribution from the $\pi i_{13/2}$ orbital. This can
be seen in Fig.~\ref{fig3} where the calculated neutron MOI of
$^{252}$No suddenly increases at $\hbar\omega\approx 0.2$ MeV,
while the proton component upbends later and contributes less
angular momentum.

The upbending behavior changes with particle numbers. It can be
seen in Fig.~\ref{fig2} that the increase of MOI becomes less
drastic with increasing neutron and proton numbers. This is
because the Fermi surface moves to be near the higher-$\Omega$
(single-particle angular momentum projection on the symmetry axis)
branches of the $\pi i_{13/2}$ and $\nu j_{15/2}$ orbitals. Such
branches tend to be deformation aligned, so that the nucleus gains
collective angular momentum more slowly.

Figure~\ref{fig2} also presents the comparison between the MOIs
calculated with and without high-order deformations. It shows that
both calculations generate almost the same MOIs for $N<152$
nuclei. However, the $N\ge 152$ nuclei have smaller MOIs at
$\hbar\omega\approx 0.25-0.35$ MeV and hence slower alignments in
the calculations with $\beta_{6,8}$, than in the calculations
without $\beta_{6,8}$. The effect comes from $\beta_6$
deformation, while the influence of $\beta_8$ deformation is
negligible.

$^{254}$No is, among others, the nucleus influenced the most by
$\beta_6$ deformation. The calculation restricted to include only
$\beta_{2,4}$ gives rise to an upbending at $\hbar\omega\approx
0.2$ MeV. This is in contrast to the gradual increase of MOI until
$\hbar\omega\approx 0.3$ MeV in the calculation incorporating
high-order deformations, which is what is observed in experiments.
Much different from $^{254}$No, $^{252}$No has a MOI barely
affected by $\beta_6$ deformation, with an upbending observed and
also calculated to occur at $\hbar\omega\approx 0.2$ MeV. A
similar difference is shown between $^{250}$Fm and $^{252}$Fm in
our calculations (see Fig.~\ref{fig2}). Here the difference is
even more distinct. $^{250}$Fm has a more drastic alignment than
$^{252}$No, and $^{252}$Fm shows a more gentle alignment than
$^{254}$No. The two isotopes would have basically the same
alignment behaviors if the $\beta_6$ deformation is ignored. The
prediction in $^{250,252}$Fm awaits experimental confirmation. The
former already has data available around the upbending frequency,
but the latter is poorly known in spectroscopy measurements
besides the $2_1^{+}$ state.

It has been demonstrated
that $\beta_6$ deformation leads to enhanced deformed shell gaps
at $Z=100$ and $N=152$~\cite{PatPLB91,LiuPRC11}. The enlargement
of the deformed shell gaps is accompanied by a lowering in energy
of the $\pi i_{13/2}$ and $\nu j_{15/2}$ orbitals below the gaps.
It is likely that the shift of the high-$j$ intruder orbitals by
$\beta_6$ deformation results in the slow alignment of $^{254}$No.
In Fig.~\ref{fig3}, we display the proton and neutron components
of the calculated MOIs for $^{252,254}$No. The comparison
indicates that the slow alignment of $^{254}$No mainly originates
from the influence of $\beta_6$ deformation on the neutron MOI.
This is because the shell gap at $N=152$ is more enhanced than at
$Z=100$ and the neutron intruder orbital has a higher $j$ value
than the proton one.

\begin{figure}
\includegraphics[scale=0.45]{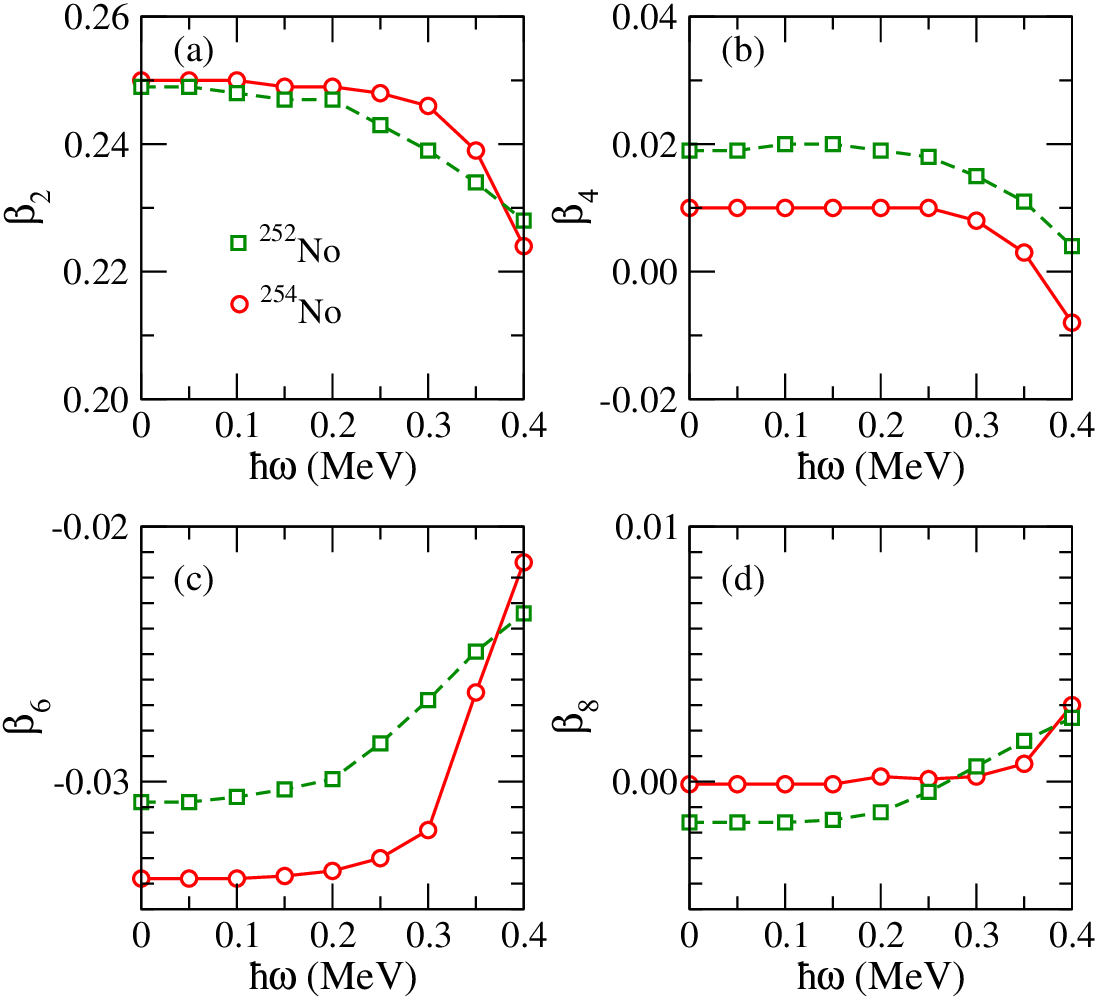}
\caption{\label{fig4}(Color online) Calculated $\beta_2$ (a),
$\beta_4$ (b), $\beta_6$ (c), and $\beta_8$ (d) deformations
versus rotational frequency for $^{252,254}$No.}
\end{figure}

The deformations of $^{252,254}$No are calculated to change with
increasing rotational frequency, as shown in Fig.~\ref{fig4}. The
deformation changes are intimately related to the rotation
alignments of the $\pi i_{13/2}$ and $\nu j_{15/2}$ orbitals. For
$^{254}$No, the $\beta_{2,4,6,8}$ deformations all keep almost
constant at first and then sharply change at $\hbar\omega\approx
0.3$ MeV. In contrast, the large changes for $^{252}$No take place
at $\hbar\omega\approx 0.2$ MeV.

In summary, TRS calculations extended to include $\beta_{6,8}$
deformations have been performed to investigate collective
rotations in transfermium nuclei. The calculated MOIs agree
satisfactorily with available data. In particular, our
calculations reproduce well the slow alignment observed in
$^{254}$No and the fast alignment observed in $^{252}$No. The underlying mechanism responsible for the
difference between $^{252}$No and $^{254}$No is found for the first time to lie in
$\beta_6$ deformation that lowers the energies of the $\nu
j_{15/2}$ intruder orbitals below the $N=152$ gap. A more distinct
difference is predicted for the $^{250,252}$Fm isotopes. Our
calculations indicate that $\beta_6$ deformation plays an vital
role around $N=152$ and $Z=100$. The inclusion of $\beta_6$
deformation results in more realistic single-particle levels
and hence has remarkable influence on
not only the multi-quasiparticle states~\cite{LiuPRC11} but also
the rotational spectra. The present work establishes a consistent
relation between high-order deformations, MOIs and excitation
energies of multi-quasiparticle states.

This work was supported by the Scientific Research Supporting Plan
of Xi'an Jiaotong University under Grant No. 08142021; the
National Natural Science Foundation of China under Grant No.
10975006; and the UK STFC and AWE plc.

\end{document}